\documentclass[conference,a4paper]{IEEEtran}
\usepackage{multirow}
\usepackage{booktabs}
\usepackage{amsfonts}
\usepackage{placeins}
\usepackage{array}
\usepackage{amsmath,cases,amssymb}
\usepackage{amsmath}
\usepackage{graphicx}
\usepackage{cite}
\usepackage{subfigure}
\usepackage{epsfig}
\usepackage{todonotes}
\usepackage{url}
\usepackage{algorithm}
\usepackage{algorithmic}
\usepackage{epstopdf}
\usepackage{balance}
\usepackage{color}
\usepackage{algorithm}
\usepackage{hyperref}
\usepackage{scalerel}

\DeclareGraphicsExtensions{.eps,.pdf}

\newcommand{\bseq}{\begin{subequations}}
\newcommand{\eseq}{\end{subequations}}
\newcommand{\baln}{\begin{align}}
\newcommand{\ealn}{\end{align}}
\newcommand{\balnd}{\begin{aligned}}
\newcommand{\ealnd}{\end{aligned}}
\newcommand{\beq}{\begin{equation}}
\newcommand{\eeq}{\end{equation}}
\newcommand{\beqn}{\begin{eqnarray}}
\newcommand{\eeqn}{\end{eqnarray}}
\newcommand{\beqno}{\begin{eqnarray*}}
\newcommand{\eeqno}{\end{eqnarray*}}
\newcommand{\bma}{\begin{displaymath}}
\newcommand{\ema}{\end{displaymath}}
\newcommand{\bnu}{\begin{enumerate}}
\newcommand{\enu}{\end{enumerate}}
\newcommand{\bce}{\begin{center}}
\newcommand{\ece}{\end{center}}
\newcommand{\btb}{\begin{tabular}}
\newcommand{\etb}{\end{tabular}}
\newcommand{\bIEEEeq}{\begin{IEEEeqnarray}}
\newcommand{\eIEEEeq}{\end{IEEEeqnarray}}

\allowdisplaybreaks

\newtheorem{theorem}{\bf Theorem}

\newtheorem{remark}{\bf Remark}

\newcommand{\st}{{\mathrm{s.t.}}}

\newcommand{\calK}{{\mathcal {K}}}
\newcommand{\calL}{{\mathcal {L}}}
\newcommand{\calM}{{\mathcal {M}}}
\newcommand{\calN}{{\mathcal {N}}}

\newcommand{\bh}{\mathbf {h}}
\newcommand{\bg}{\mathbf {g}}

\newcommand{\balpha}{\boldsymbol{\alpha}}

\newcommand{\bmu}{\boldsymbol{\mu}}

\newcommand{\BS}{\mathtt{BS}}
\newcommand{\BSn}{\mathtt{BS}_n}

\newcommand{\UE}{\mathtt{UE}}
\newcommand{\SUE}{\mathtt{SUE}}
\newcommand{\LEO}{\mathtt{LEO}}

\newcommand{\bp}{\mathbf{p}}
\newcommand{\bP}{\mathbf{P}}

\newcommand{\bW}{\mathbf{W}}

\makeatletter
\newcommand{\linebreakand}{%
\end{@IEEEauthorhalign}
\hfill\mbox{}\par
\mbox{}\hfill\begin{@IEEEauthorhalign}
}
\makeatother

\begin{document}

\bstctlcite{IEEEexample:BSTcontrol}
\title{LEO-to-User Assignment and Resource Allocation for Uplink Transmit Power Minimization}

\author{\IEEEauthorblockN{Hung Nguyen-Kha}
\IEEEauthorblockA{\textit{SnT, University of Luxembourg} \\
khahung.nguyen@uni.lu}\vspace{-0.2cm}
\and
\IEEEauthorblockN{Vu Nguyen Ha}
\IEEEauthorblockA{\textit{SnT, University of Luxembourg} \\
vu-nguyen.ha@uni.lu}\vspace{-0.2cm}
\and
\and\IEEEauthorblockN{Eva Lagunas}
\IEEEauthorblockA{\textit{SnT, University of Luxembourg} \\
eva.lagunas@uni.lu}\vspace{-0.2cm}
\linebreakand 
\IEEEauthorblockN{Symeon Chatzinotas}
\IEEEauthorblockA{\textit{SnT, University of Luxembourg} \\
symeon.chatzinotas@uni.lu}\vspace{-0.8cm}
\and
\IEEEauthorblockN{Joel Grotz}
\IEEEauthorblockA{\textit{SES S.A., Luxembourg} \\
joel.grotz@ses.com}\vspace{-0.8cm}
}

\maketitle

\begin{abstract}
This paper aims to develop satellite--user association and resource allocation mechanisms to minimize the total transmit power for integrated terrestrial and non-terrestrial networks wherein a constellation of LEO satellites provides the radio access services to both terrestrial base stations (BSs) and the satellite-enabled users (SUEs). 
In this work, beside maintaining the traditional SatCom connection for SUEs, the LEO satellites provide backhaul links to the BSs to upload the data received from their ground customers.
Taking the individual SUE traffic demands and the aggregated BS demands, we formulate a mixed integer programming which consists of the binary variables due to satellite association selection, power control and bandwidth allocation related variables. To cope with this challenging problem, an iterative optimization-based algorithm is proposed 
by relaxing the binary components and alternating updating all variables. 
A greedy mechanism is also presented for comparison purpose. Then, numerical results are presented to confirm the effectiveness of our proposed algorithms.
\end{abstract}
\begin{IEEEkeywords}
Integrated terrestrial and non-terrestrial networks, LEO constellation, resource allocation, satellite association, power minimization.	
\end{IEEEkeywords}

\section{Introduction}
Recently, the rapid growth of number of devices, broadband traffic, as well as the new applications in the internet of things (IoT) era has challenged the terrestrial network (TN) operators \cite{6G_UseCase_Technologies,DuyVu_VTC2019,VuHa_TVT2016,NTN_6G_challenges_opportunities_IEEENetwork}. 
Although fifth generation (5G) new radio cellular networks have been developed and deployed, which can bring several advantages, such as, high data rate, high energy efficiency, and low latency, the critical issue of ubiquitous coverage at the rural and city-edge areas has not been addressed efficiently due the high cost of backhaul-link and transportation network implementation \cite{DuyVu_VTC2019,VuHa_TVT2016, NTN_6G_challenges_opportunities_IEEENetwork, Satcom_survey_challenges}. 

To overcome these challenges, the low-earth-orbit (LEO) satellite and TN integration has been considered as a promising solution \cite{NTN_6G_challenges_opportunities_IEEENetwork,Satcom_survey_challenges,Satcom_at_milimeterwave_ICNC,VuHaGC2022,VuHa_TWC22,VuHa_VTCFall22,VuHa_GCWks22}. Thanks to inter satellite links and low orbit altitude, the LEO constellation can not only supply the global connectivity with wide coverage but also provide sufficiently-low-latency connection and high-data-rate services. 
Therefore, the LEO satellite constellation can contribute the high-speed and reliable backhaul connection between isolated cells and core network \cite{Satcom_survey_challenges, Use_cases_5G_satllite}. 
However, enhancing the LEO connectivity as the backhaul links for terrestrial BSs has also required to re-solve several traditional technical issues such as SUE/BS association, power control, and bandwidth allocation for the new integrated terrestrial and non-terrestrial systems \cite{NTN_6G_challenges_opportunities_IEEENetwork}.

Regarding the satellite and SUEs/BSs association, some satellite-user association schemes have been proposed in the literature \cite{Handover_MaxWeight_BipartiteBraph, Handover_LEO_PotentialGame, LEO_TN_2tier_greedy-association}.
The satellite selection based on the channel strength has been proposed for ground users in \cite{GlobeW_5GNR_LEO_performance}; however, the channel gain may vary too frequently in the practical systems which may degrade network performance. Furthermore, this work only focuses on the satellite selection while the data demand and the limitation of satellite capacity are not considered.
In \cite{Handover_MaxWeight_BipartiteBraph}, the authors proposed a satellite-gateway association strategy using the graph-theory approach, where a bipartite graph is established with the weight set from the channel gain and the coverage of satellites. Accordingly, the association solution is obtained by finding the maximum weighted matching. While the satellite selection problem is considered as a potential game in \cite{Handover_LEO_PotentialGame}. 
In \cite{LEO_TN_2tier_greedy-association}, the LEO satellite is deployed to assist the terrestrial network. This work aims to optimize the user association and transmission power with the fixed bandwidth allocation to maximize the total network throughput. 
To the best of our knowledge, the problem of joint LEO satellite association, resource allocation under constraints of data demand and load balancing has not been considered in the previous related works, that motivates us to investigate these issues.

In this paper, we aim to minimize the uplink transmit power for the integrated LEO satellite and TN system serving both satellite and NT users. 
In this system, the LEO satellites can provide the backhaul link for the cellular BSs and the radio access service for the SUE simultaneously. 
Regarding the demand for satellite and terrestrial users, we formulate the power minimization problem which jointly design the LEO association, transmit power, and bandwidth allocation as an NP-hard mixed interger programming (MIP).
Particularly, the optimization problem is difficult to solve directly owing to the coupling of binary and continuous variables, and the non-convexity of rate and objective functions. Hence, we propose an iterative algorithm based on alternating optimization method to cope with it. 
The numerical results demonstrate the convergence and effectiveness of the proposed algorithm.

The rest of the paper is organized as follows. Section~\ref{Sec:sysmodel} shows the system model and problem formulation. Section~\ref{Sec:Solution} describes our proposed iterative algorithm and greedy-based algorithm, respectively. Section~\ref{Sec:simul} presents the numerical results which is followed by the conclusion given in Section~\ref{Sec:Concl} .

\section{System Model and Problem Formulation}
\label{Sec:sysmodel}

\begin{figure}[!h]
\centering
\includegraphics[width=0.8\linewidth]{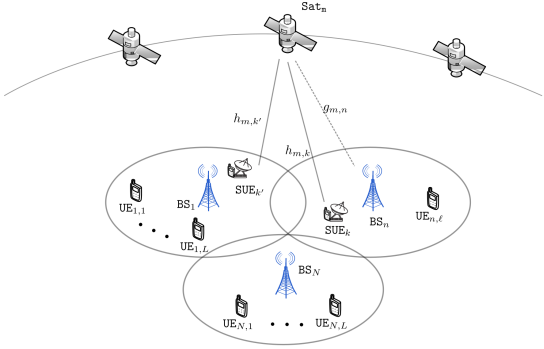}
\caption{An example of ISTN communication system.}
\label{fig:LEO_BS_SUE}
\end{figure}

We consider the uplink transmission of an integrated satellite-terrestrial network (ISTN) communication system  consisting of $M$ LEO satellites, $N$ ground base stations with their users, and $K$ SUEs as described in Fig.~\ref{fig:LEO_BS_SUE}. 
Let $\calM$, $ \calN$, $\calL_n$ and $ \calK$ denote the sets of LEO satellites, BSs, users (UEs) in cell $n$ served by BS $n$, and SUEs, respectively.
In this integrated system, together with providing the radio access service to SUEs as in traditional SATCOMs, the LEO satellites also provide backhaul service to BS to offload data from their users to the core-network. 
For convenience, we define $m-$th LEO satellite, $n-$th BS, $ \ell- $th UE of BS $n$, and $k-$th SUE by $\LEO_{m}, \BSn, \UE_{n,\ell}$ and $\SUE_k$, respectively.  
The channel gain of links $\LEO_{m}-\SUE_k$ and $\LEO_{m}-\BSn$ are modeled as 
\bIEEEeq{ll}
h_{m,k}=\scaleobj{.9}{\frac{G^\LEO G^\SUE \psi(\vartheta_{m,k})}{ \mathtt{PL}_{m,k}}} \text{ and }
g_{m,n}=\scaleobj{.9}{\frac{G^\LEO G^\BS \psi(\vartheta_{m,n})}{ \mathtt{PL}_{m,n}}}, 
\eIEEEeq
where $\mathtt{PL}_{m,i} = (4 \pi f_c d_{m,i} / c)^2$ and $d_{m,i}$ denote the free space path loss and the distance between LEO satellite $m$ and SUE/BS $i$; $G^\LEO,G^\SUE$ and $G^\BS$ are the maximum antenna gains of the LEO satellites, SUEs and BSs; $\vartheta_{m,k}$ and $\vartheta_{m,n}$ are the boresight angle from $\LEO_{m}$ to $\SUE_k$ and $\BS_n$, respectively. 
Additionally, taking into account the channel in \cite{3gpp.38.811}, 
$\psi(\vartheta)$ in the channel gain formulas represents the beam pattern function which is expressed as 
\bIEEEeq{ll}
    \psi(\vartheta)=
    \begin{cases}
        1 &, \vartheta=0, \\
        4 |\frac{J_1(k a \sin{\vartheta})}{k a \sin{\vartheta}}| &, \vartheta \neq 0,
    \end{cases} \nonumber
\eIEEEeq
where $k=2 \pi f_c / c$, $a, f_c$, and $c$ are the antenna aperture radius, operation frequency and light speed, respectively.
\subsection{Communication between SUE and Satellite}
Regarding the LEO-SUE links, one assumes that the radio-access service according to every LEO is available to all SUEs and BSs located inside its corresponding coverage.
Then, each SUE can associate with an LEO satellite for data-transmission purpose depended on its location, its transmission budget, and the available radio resource at the satellite. 
To perform the LEO-SUE association, we introduce a new variable $ \balpha \triangleq [\alpha_{m,k}]_{\forall (m,k) \in (\calM \times \calK)} $ as
\begin{align} \label{eq: BS-SUE association}
\alpha_{m,k}= 
\begin{cases}
	1, &  \SUE_{k} \text{ is served by } \LEO_{m}, \\
	0 ,&  \text{otherwise}.
\end{cases}
\end{align}
Due to the directional antennas employed in SATCOM, one presumes that each SUE can be served by at most one LEO satellite selected from covering ones, which yields the following constraint, 
\beq
(C1): \quad \scaleobj{.8}{\sum_{\forall m \in \calM}} \alpha_{m,k} \leq 1, \forall k \in \calK.
\eeq
Once $\SUE_{k}$ is served by $\LEO_{m}$, 
let $W^\SUE_{k}$ be the bandwidth allocated to $\SUE_{k}$ and $p_{m,k}$ indicate the transmission power of this user.
The orthogonal bandwidth assignment is assumed in this work based on which the
Signal-to-noise ratio (SNR) of $\SUE_{k}$ can be written as
\begin{IEEEeqnarray}{ll} \label{eq: SINR SUE_ell at BSn SC s}
\gamma^\SUE_{m,k} = \frac{ p_{m,k} h_{m,k} }{ \sigma_m W^\SUE_k},
\end{IEEEeqnarray}
where $\sigma_m$ is the noise power density per Hz at $ \LEO_{m} $. 
Then, the achievable rate of $ \SUE_{k} $ at $\LEO_{m}$ can be expressed as
\begin{IEEEeqnarray}{ll} \label{eq: rate SUE_ell at LEOm}
R_{m,k}^\SUE \!\! = W^\SUE_{k} \log_2(1 \!\! + \!\! \gamma^\SUE_{m,k} ) = W^\SUE_{k} \log_2 \!\! \left( \!\! 1 \!\! + \!\! \frac{ p_{m,k} h_{m,k} }{ \sigma_m W^\SUE_k} \!\! \right).
\end{IEEEeqnarray}
Taking into account the LEO association decision, the achievable transmission rate of $ \SUE_{k} $ can be described as
\beq \label{eq: rate UEk}
R_k^\SUE (\bp,\bW^\SUE,\balpha) = \scaleobj{.8}{\sum_{\forall m \in \calM}} \alpha_{m,k} R_{m,k}^\SUE,
\eeq
where $\bp \triangleq [p_{m,k}]_{\forall m,k}$ and $\bW^\SUE \triangleq [W^{\mathtt{SUE}}_{k}]_{\forall k}$.
Regarding the communication rate demand at each SUE, the following constraint is introduced,
\beq
(C2): \quad R^\SUE_{k} (\bp,\bW^\SUE,\balpha) \geq \bar{R}^\SUE_{k}, \quad \forall k \in \calK,
\eeq
in which $\bar{R}^\SUE_k $ indicates the required transmission rate of $\SUE_k$.
\subsection{Data Offloading from BS to LEO}
In this system, we assume that the uplink data transmission of terrestrial UEs in one cell is gathered by the corresponding BS before being transferred to the core network
through the satellite return links.
Let $\bar{R}^\UE_{n,\ell}$ be the required data rates of user $\ell$ in cell $n$, i.e. $ \UE_{n,\ell}$. 
Then, the required offloaded data from $\BSn$ can be defined as $\bar{R}^{\mathtt{BS}}_n = \sum_{\forall \ell \in \calL_n} \bar{R}^\UE_{n,\ell}$.
Like SUEs, each BS will select a LEO satellite to associate with for data offloading. Denote $ \bmu \triangleq [\mu_{m,n}]_{\forall (m,n) \in (\calM \times \calN)} $ as the association variable between BSs and LEOs which is determined as
\begin{align} \label{eq: BS-LEO association}
\mu_{m,n}= 
\begin{cases}
	1, &  \BSn \text{ is served by } \LEO_{m}, \\
	0 ,&  \text{otherwise}.
\end{cases}
\end{align}
Note that each BS can be served by at most one LEO satellite, similar to $(C1)$, this is also casted by the following constraint,
\beq
(C3): \quad \scaleobj{.8}{\sum_{\forall m \in \calM}} \mu_{m,n} \leq 1, \forall n \in \calN, 
\eeq
Let $W^\BS_n$ be the bandwidth of the return channel assigned to $\BSn$. Then, once $\BSn$ is served by $\LEO_{m}$, the SNR of $ \BSn $ transmission signal received at $\LEO_{m}$ can be expressed as
\begin{IEEEeqnarray}{ll} \label{eq: SNR BSn at LEOm}
\gamma^\BS_{m,n}  = \scaleobj{.8}{\frac{ P_{m,n} g_{m,n}}{\sigma_m W^\BS_n }},
\end{IEEEeqnarray}
where $P_{m,n} $ denotes the transmit power intended for $\LEO_{m}$ at $\BS_n$.
Hence, the achievable offloading rate of $\BSn$ at $\LEO_{m}$ is written as
\begin{IEEEeqnarray}{ll} \label{eq: rate BSn at LEOm}
R_{m,n}^\BS \!\! = \!\! W^\BS_n \log_2(1 \!\! + \!\! \gamma^\BS_{m,n}) = W^\BS_n \log_2 \!\! \scaleobj{.8}{\left( \!\! 1 \!\! + \!\! \frac{ P_{m,n} g_{m,n} }{ \sigma_m W^\BS_n} \!\! \right)}.
\end{IEEEeqnarray}
Considering the LEO association selection of $ \BSn $, it offloading rate can be summarized as
\beq
R_n^\BS (\bP,\bW^\BS,\bmu) = \scaleobj{.8}{\sum_{\forall m \in \calM}} \mu_{m,n} R_{m,n}^\BS,
\eeq
where $\bP \triangleq [P_{m,n} ]_{\forall m,n}$ and and $\bW^\BS \triangleq [W^{\mathtt{BS}}_{n}]_{\forall n}$.
In order to successfully forward all the data from UEs associated to $\BSn$ to the core network, the following condition must be hold,
\beq
(C4): \quad R_n^\BS (\bP,\bW^\BS,\bmu) \geq \bar{R}^\BS_n
, \forall n \in \calN.
\eeq

\subsection{Problem Formulation} 
In this work, we aim to minimize the total transmit power of all BSs and SUEs while satisfying all users' transmission-rate demands. 
The design objective can be mathematically formulated as following power minimization problem, 
\bIEEEeq{cl}\label{eq: Min power}
\min_{\scaleobj{.8}{\substack{\bp,\bP,\bW^\SUE,\\ \bW^\BS,\balpha,\bmu}}}  \quad &  {\scaleobj{.8}{\sum \limits_{\substack{\forall m \in \calM \\ \forall n \in \calK}}}} \alpha_{m,k} p_{m,k} + \scaleobj{.8}{\sum \limits_{\substack{\forall m \in \calM \\ \forall n \in \calN}}} \mu_{m,n} P_{m,n}	\\
\st     & \text{constraints } (C1)-(C4), \nonumber \\
& (C5): \quad \scaleobj{.8}{\sum_{\forall m \in \calM}} \alpha_{m,k} p_{m,k} \leq p_k^{\max}, \forall k \in \calK, \nonumber \\
& (C6): \quad \scaleobj{.8}{\sum_{\forall m \in \calM}} \mu_{m,n} P_{m,n} \leq P_n^{\max}, \forall n \in \calN, \nonumber \\
& (C7): \quad  \scaleobj{.8}{\sum_{\forall k \in \calK}} \alpha_{m,k} W^\SUE_{k} \nonumber \\
& \quad\quad\quad\quad + \scaleobj{.8}{\sum_{\forall n \in \calN}} \mu_{m,n} W^\BS_n  \leq W^\LEO_{m}, \quad \forall m \in \calM, \nonumber \\
& (C8): \quad \alpha_{m,k}, \mu_{m,n} \in \{0,1\}, \forall (m,n,k), \nonumber  
\eIEEEeq
in which constraints $(C5)$ and $(C6)$ represent the limitation on maximum transmit power of SUEs and BSs. While constraint $(C7)$ implies that the total allocated bandwidth for serving BSs and SUEs at each LEO satellite is lower than a pre-determined amount of available bandwidth.

\section{Proposed Solution Approaches} \label{Sec:Solution}
As can be observed, the original optimization problem \eqref{eq: Min power} is non-convex and it is difficult to be solved directly owing to the coupling between binary association variables and power and bandwidth allocation ones. To address this critical issue and develop an efficient mechanism for solving problem \eqref{eq: Min power}, we first decompose it into two sub-problems. 
In particular, the first one aims to optimize the power and bandwidth allocation variables $\bP, \bp, \bW^\SUE$ and $\bW^\BS$ for given LEO association decision. While the remaining one focuses on determining the association variables $\balpha$ and $\bmu$ to reduce the total transmission power for given values of $\bP, \bp, \bW^\SUE$ and $\bW^\BS$.
Then, the solution of \eqref{eq: Min power} will be defined by alternatively and iteratively solving these two sub-problems.

\subsection{Power and bandwidth allocation optimization}
At the iteration $i$, once $(\balpha,\bmu)$ are fixed at $(\balpha^{(i)},\bmu^{(i)})$, the corresponding power and bandwidth allocation, i.e., $\bP, \bp, \bW^\SUE$ and $\bW^\BS$ can be optimized by addressing the following optimization problem, 
\bIEEEeq{cl}\label{eq:MinPower, P BW}
\min_{\scaleobj{.8}{\substack{\bp,\bP,\\ \bW^\SUE, \bW^\BS}}}  \quad &  {\scaleobj{.8}{\sum_{\substack{\forall m \in \calM \\ \forall k \in \calK}}}} \bar{\alpha}_{m,k} p_{m,k}	+ \scaleobj{.8}{\sum_{\substack{\forall m \in \calM \\ \forall n \in \calN}}} \bar{\mu}_{m,n} P_{m,n}	\\
\st
& (C2.1): \quad  \bar{R}^\SUE_{k} \leq R^\SUE_{k} (\bp,\bW^\SUE,\bar{\balpha}), \quad \forall k \in \calK, \nonumber \\
& (C4.1): \quad \scaleobj{.8}{\sum_{\forall \ell \in \calL_n}} \bar{R}^\UE_{n,\ell} \leq R_n^\BS (\bP,\bW^\BS,\bar{\bmu}), \forall n \in \calN, \nonumber \\
& (C5.1): \quad \scaleobj{.8}{\sum_{\forall m \in \calM}} \bar{\alpha}_{m,k} p_{m,k} \leq p_k^{\max}, \forall k \in \calK, \nonumber \\
& (C6.1): \quad \scaleobj{.8}{\sum_{\forall m \in \calM}} \bar{\mu}_{m,n} P_{m,n} \leq P_n^{\max}, \forall n \in \calN, \nonumber \\\
& (C7.1): \quad  \scaleobj{.8}{\sum_{\forall k \in \calK}} \bar{\alpha}_{m,k} W^\SUE_{k} \nonumber \\
& \quad\quad\quad\quad +  \scaleobj{.8}{\sum_{\forall n \in \calN}} \bar{\mu}_{m,n} W^\BS_n \leq W^\LEO_{m}, \quad \forall m \in \calM, \nonumber 
\eIEEEeq
where $(\bar{\balpha},\bar{\bmu})=(\balpha^{(i)},\bmu^{(i)})$, and $(C2.1),(C4.1)-(C7.1)$ are updated from $(C2),(C4)-(C7)$ with a given value of association variable. To address this problem, we first characterize its convexity in the following theorem.
\begin{theorem} \label{P01:thm_01}
Problem \eqref{eq:MinPower, P BW} is convex.
\end{theorem}
\begin{IEEEproof}
It can be seen that rate functions $R^\SUE_{k} (\bp,\bW^\SUE,\bar{\balpha})$ and $R_n^\BS (\bP,\bW^\BS,\bar{\bmu})$ are concave with the fixed value $(\bar{\balpha},\bar{\bmu})$, resulting in that constraints $(C4.1)$ and $(C5.1)$ are convex. In addition, constraints $(C6.1),(C7.1)$ and the objective function in \eqref{eq:MinPower, P BW} are linear with fixed association variables. Therefore, problem \eqref{eq:MinPower, P BW} is convex.
\end{IEEEproof}

Thanks to \textbf{Theorem \ref{P01:thm_01}}, one can note that the optimal  solution of problem \eqref{eq:MinPower, P BW} can be obtained by utilizing some standard optimization tools, e.g., Gurobi, Mosek, GAMS, CVX \cite{cvx}.

\subsection{LEO Association Update: Relaxation and Projection}
This section considers the LEO association decision at BSs and SUEs in the iteration $i$ when transmit power and bandwidth allocation are fixed as $(\bp^{(i+1)},\bP^{(i+1)},(\bW^\SUE)^{(i+1)}, (\bW^\BS)^{(i+1)})$. Accordingly, the LEO association optimization problem at iteration $(i+1)$ can be written as
\bIEEEeq{cl}\label{eq: Min power, binary variable}
\min_{\bmu,\balpha}  \quad & {\scaleobj{.8}{\sum_{\substack{\forall m \in \calM \\ \forall n \in \calN}}}} \mu_{m,n} \bar{P}_{m,n} + \scaleobj{.8}{\sum_{\substack{\forall m \in \calM \\ \forall k \in \calK}}} \alpha_{m,k} \bar{p}_{m,k}		\\
\st     & \text{constraints } (C1),(C3), \text{ and } (C8), \nonumber \\
& (C2.2): \quad  \bar{R}^\SUE_{k} \leq R^\SUE_{k} (\bar{\bp},\bar{\bW}^\SUE,\balpha), \quad \forall k \in \calK, \nonumber \\
& (C4.2): \quad \scaleobj{.8}{\sum_{\forall \ell \in \calL_n}} \bar{R}^\UE_{n,\ell} \leq R_n^\BS (\bar{\bP},\bar{\bW}^\BS,\bmu), \forall n \in \calN, \nonumber \\
& (C5.2): \quad \scaleobj{.8}{\sum_{\forall m \in \calM}} \mu_{m,n} \bar{P}_{m,n} \leq P_n^{\max}, \forall n \in \calN, \nonumber \\\
& (C6.2): \quad \scaleobj{.8}{\sum_{\forall m \in \calM}} \alpha_{m,k} \bar{p}_{m,k} \leq p_k^{\max}, \forall k \in \calK, \nonumber \\
& (C7.2): \quad \scaleobj{.8}{\sum_{\forall n \in \calN}} \mu_{m,n} \bar{W}^\BS_n \nonumber \\
& \quad\quad\quad\quad + \scaleobj{.8}{\sum_{\forall k \in \calK}} \alpha_{m,k} \bar{W}^\SUE_{k} \leq W^\LEO_{m}, \quad \forall m \in \calM, \nonumber 
\eIEEEeq
where $(\bar{\bp},\bar{\bP},\bar{\bW}^\SUE, \bar{\bW}^\BS)=\left(\bp^{(i+1)},\bP^{(i+1)},(\bW^\SUE)^{(i+1)}\right.$ $\left., (\bW^\BS)^{(i+1)}\right)$. Constraints $(C2.2),(C4.2)-(C7.2)$ are rewritten from $(C2),(C4)-(C7)$ with fixed values of power and bandwidth allocation. As can be observed, problem \eqref{eq: Min power, binary variable} is an integer linear programming which can be solved efficiently by utilizing some standard optimization tools, e.g., Matlab ('intlinprog' function), Gurobi, Mosek, GAMS, CVX \cite{cvx}.

\begin{remark}
It is worth noting that the outcomes of solving  \eqref{eq: Min power, binary variable} are the binary values.
If these binary numbers are utilized to update power and bandwidth allocation as in problem \eqref{eq:MinPower, P BW}, the solution in the following iteration may be trapped at a local optimal point which cannot be improved \cite{Sanjabi_TSP14,VuHa_TWC18,VuHa_Access17}.
\end{remark}

To avoid the case where sub-problem \eqref{eq:MinPower, P BW} is strapped with strictly binary, a relaxation and projection approach similar to what introduced in \cite{Sanjabi_TSP14,VuHa_TWC18,VuHa_Access17} is employed. In particular, after obtaining the binary solution $(\balpha^\star,\bmu^\star)$ for \eqref{eq: Min power, binary variable}, to avoid the case where sub-problem \eqref{eq:MinPower, P BW} is strapped with strictly binary feasible point $(\balpha^{(i)},\bmu^{(i)})$, we use the following update
\beq \label{eq: update binary variable}
    (\balpha^{(i+1)},\bmu^{(i+1)}) = (1-\varrho) (\balpha^{(i)},\bmu^{(i)}) + \varrho (\balpha^\star,\bmu^\star),
\eeq
where $0 < \varrho < 1 $ is the update rate.
By alternatively and iteratively optimizing the transmit power and bandwidth allocation $(\bp,\bP,\bW^\SUE,\bW^\BS)$ by solving \eqref{eq:MinPower, P BW} and updating $(\balpha,\bmu)$ as in \eqref{eq: update binary variable}, problem \eqref{eq:MinPower, P BW} can be solved effectively. The proposed algorithm is summarized in Algorithm \ref{alg: proposed}.

\begin{algorithm}[t]
\footnotesize
	\begin{algorithmic}[1]
		\protect\caption{Proposed Algorithm to Solve Problem \eqref{eq: Min power}}
		\label{alg: proposed}
		\long\def\algorithmicrequire{\textbf{Phase 1:}}
		\REQUIRE
		\STATE{\textbf{Initialization:}} \\
        \STATE Set $i:=0$ and generate an initial point $(\balpha^{(0)},\bmu^{(0)})$.\\
		\REPEAT
		\STATE Solve \eqref{eq:MinPower, P BW} to obtain $(\bp^\star,\bP^\star,(\bW^\SUE)^\star,(\bW^\BS)^\star)$.
		\STATE Update $(\bp^{(i+1)},\bP^{(i+1)},(\bW^\SUE)^{(i+1)},(\bW^\BS)^{(i+1)})=(\bp^\star,\bP^\star,(\bW^\SUE)^\star,(\bW^\BS)^\star)$.
		\STATE Solve integer linear problem \eqref{eq: Min power, binary variable} to obtain $(\balpha^\star,\bmu^\star)$
		\STATE Update $(\balpha^{(i+1)},\bmu^{(i+1)})$ using \eqref{eq: update binary variable}
		\UNTIL Convergence
		\STATE \textbf{Output-1:} The optimal solution $(\bp^\star,\bP^\star,(\bW^\SUE)^\star,(\bW^\BS)^\star,\balpha^\star,\bmu^\star)$ with the continuous value of $(\balpha^\star,\bmu^\star)$.
		\long\def\algorithmicrequire{\textbf{Phase 2:}}
		\REQUIRE
		\STATE Round $(\balpha^\star,\bmu^\star)$ to recovery the binary value.\\
		\STATE Run step 4 with restored association variables to obtain exact solution $(\bp^\star,\bP^\star,(\bW^\SUE)^\star,(\bW^\BS)^\star)$.\\
		\STATE \textbf{Output-2:} The optimal solution $(\bp^\star,\bP^\star,(\bW^\SUE)^\star,(\bW^\BS)^\star,\balpha^\star,\bmu^\star)$ for problem \eqref{eq: Min power}.
\end{algorithmic} 
\normalsize
\end{algorithm}

\subsection{Greedy-Based Solution}
For comparison purpose, this section introduce an greedy algorithm for solving problem \eqref{eq:MinPower, P BW} which is summarized in Algorithm \ref{alg: greedy channel gain}.
Following this low-complexity approach, each LEO satellite is assumed to serve at most $ \lfloor N / M + 0.5 \rfloor$ BSs and $ \lfloor K / M + 0.5 \rfloor$ SUEs to ensure that all LEO satellites jointly serve BSs and SUEs. The pairs of LEO satellite and BS, and SUE with the best channel gain are associated until all BSs and SUEs are served. 
Subsequently, each LEO satellite allocates uniformly bandwidth for all served SUEs and BSs. The greedy-based algorithm is summarized in Algorithm~\ref{alg: greedy channel gain}. For the convenience, we use the Matlab notation for the index of channel gain matrices $\bg$ and $\bh$.

\begin{algorithm}[t]
\footnotesize
\begin{algorithmic}[1]
    \protect\caption{Greedy Algorithm}
    \label{alg: greedy channel gain}
    \STATE {\textbf{Input:} Channel gain matrices $ \bh $ and $ \bg $}
    \STATE Set the number of available connection from $ \LEO_{m} $ to BSs and SUEs: $ N^{\BS,\mathtt{a}}_{m} := \lfloor N / M + 0.5 \rfloor $ and $ N^{\SUE,\mathtt{a}}_{m} := \lfloor K / M + 0.5 \rfloor$ 
    \REPEAT
    \STATE Find the index of the maximum element in $\bg$: $(m^{\max},n^{\max})$
    \IF {$N^{\BS,\mathtt{a}}_{m} > 0$}
        \STATE Update $\mu_{m^{\max},n^{\max}} = 1, N^{\BS,\mathtt{a}}_{m} = N^{\BS,\mathtt{a}}_{m} - 1$ and \\ $\bg(:,n^{\max}) = \mathbf{0}$
    \ELSE
        \STATE Update $\bg(m^{\max},:) := \mathbf{0}$
    \ENDIF
    \UNTIL All BSs are assigned 
    \REPEAT
    \STATE Find the index of the maximum element in $\bh$: $(m^{\max},k^{\max})$
    \IF {$N^{\SUE,\mathtt{a}}_{m} > 0$}
        \STATE Update $\mu_{m^{\max},k^{\max}} = 1, N^{\SUE,\mathtt{a}}_{m} = N^{\SUE,\mathtt{a}}_{m} - 1$ and \\ $\bh(:,k^{\max}) = \mathbf{0}$
    \ELSE
        \STATE Update $\bh(m^{\max},:) := \mathbf{0}$
    \ENDIF
    \UNTIL All SUEs are assigned \\
    \STATE Calculate bandwidth per user in $\LEO_{m}$: $\bar{W}^{\mathtt{per}}_m = W^{\LEO}_{m} / (\sum_{\forall k} \alpha_{m,k} + L \sum_{\forall n} \mu_{m,n} )$
    \STATE Set $W^{\SUE}_{k} := \sum_{\forall m} \alpha_{m,k} \bar{W}^{\mathtt{per}}_m $ and $W^{\BS}_{n} := L \sum_{\forall m} \mu_{m,n} \bar{W}^{\mathtt{per}}_m $
    \STATE \textbf{Output:} $\bmu, \balpha, W^{\BS}_n$ and $W^{\SUE}_k, \forall n,k$
\end{algorithmic} 
\normalsize
\end{algorithm}

\section{Numerical Results} \label{Sec:simul}

\begin{table}[t]
	\caption{Simulation Parameters}
	\label{tab:parameter}
	\centering
	\begin{tabular}{l|l}
		\hline
		Parameter & Value \\
		\hline\hline
        LEO satellite bandwidth, $W^\LEO_{m}$   &  500 MHz \\
        LEO satellite altitude                  & 340 km \\
        LEO satellite antenna gain, $G^\LEO$    & 42 dBi \\
        BS antenna gain, $G^\BS$                & 32.8 dBi \\
        SUE antenna gain, $G^\SUE$              & 10 dBi \\
        Operation frequency, $f_c$              & 27.5 GHz \\
		Noise power power density, $\sigma_m$	& -174 dBm/Hz \\
        Maximum power at SUE, $ p^{\max}_{k} $	& 20 dBW \\
		Maximum power at BS, $ P^{\max}_{n} $	& 40 dBW \\
		Number of SUEs, $ K $					& 10 \\
        Number of BSs, $ N $					& 10 \\
		Demand per user, $ \bar{R}^\SUE_{k}=\bar{R}^\UE_{n,\ell}=\bar{R} $	& 100 Mbps \\
		\hline		   				
	\end{tabular}
\end{table}

\begin{figure}
    \centering
    \includegraphics[width=0.7\linewidth]{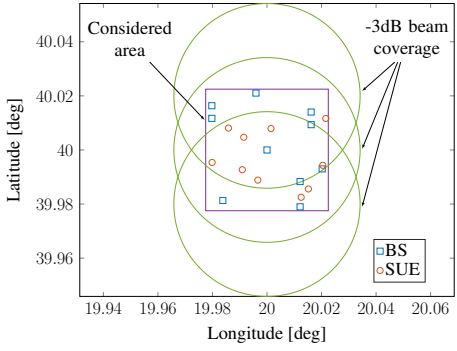}
    \caption{A simulation topology with $K=10$ SUEs, $N=10$ BSs.}
    \label{fig:Simulation_topo}
    \vspace{-2mm}
\end{figure}

In this section, the numerical results are illustrated to evaluate the performance of the proposed algorithms. The simulation parameters are summarized as in Table~\ref{tab:parameter}. In these simulation, we consider a square area of $25$~$\text{km}^2$ whose center is located at geographical coordinate $(\varphi_{0},\theta_{0}) = (40^\circ \text{N}, 20^\circ \text{E})$.
Inside this area, $K$ SUEs and $N$ BSs are randomly deployed. Herein, the number of UEs connecting to each BS is set by utilizing a Poisson distribution with a mean of $\bar{L} = 10$. Furthermore, $3$ LEO satellites ($M=3$) are assumed to cover this area which are taken from one orbit of the Walker star constellation. 
The latitude difference of two adjacent LEO satellites is $0.02^\circ$, the middle LEO satellite has the projection at $(\varphi_{0},\theta_{0})$. 
For instance, a simulation topology consisting of $K=10$ SUEs, $N=10$ BSs served by three LEO satellites is illustrated in Fig.~\ref{fig:Simulation_topo}. 

\begin{figure}
    \centering
    \includegraphics[width=0.95\linewidth]{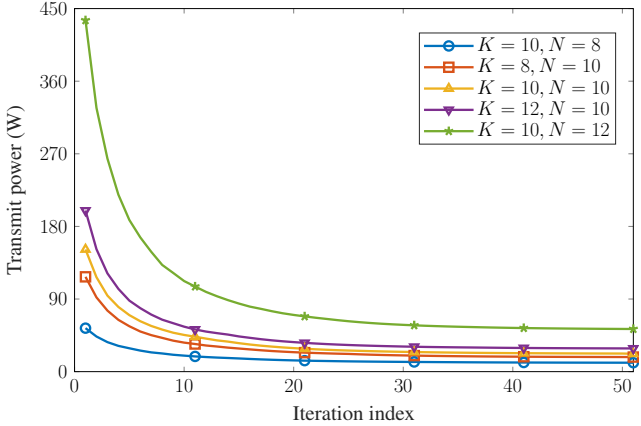}
    \caption{Convergence rate of the proposed algorithm.}
    \label{fig:P_convergence}
    \vspace{-2mm}
\end{figure}

Fig. \ref{fig:P_convergence} illustrates the convergence behavior of the proposed algorithm with different numbers of SUEs and BSs, i.e., $K=8,10,12$ and $N=8,10,12$. As can be observed, 
the total transmission power decreases and then saturates after a few number of iterations, which has confirmed the convergence of the proposed algorithm.
It can be seen that the higher number of SUEs and BSs results in a higher number of iterations for convergence implementation.
In particular, the schemes according to $(K,N) = (10, 8)$ and $(8,10)$ require about $16$ iterations for convergence while that number corresponding to $(10,10)$ and $(10,12)$ (or $(12,10)$) schemes are around $22$ and $31$, respectively. 
As expected, implementing Algorithm~\ref{alg: proposed} for a larger scaled-size system requires a higher number of iterations to converge.

\begin{figure}
    \centering
    \includegraphics[width=0.95\linewidth]{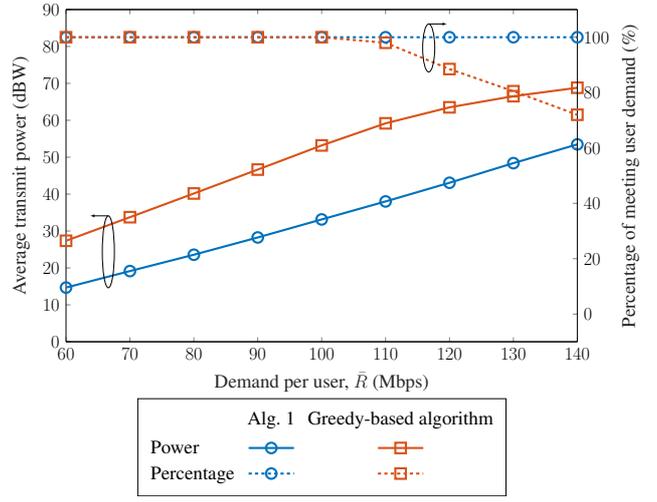}
    \caption{Average transmit power versus the demand per user.}
    \label{fig:P_QoS}
\end{figure}

The impact of demand per user $\bar{R}$ on transmit power is shown in Fig. \ref{fig:P_QoS}, where the total transmission according to two propose algorithms versus various users' demand is illustrated.
As anticipated, BSs and SUEs consume more power to meet the higher user demand for both algorithms. Especially, the transmit power returned by both mechanisms raises quickly when $\bar{R}$ increases.
In particular, the consumed-power gap between two points $\bar{R}=60$ Mbps and $\bar{R}=120$ Mbps of Greedy-based algorithm is about $36$ dB, while that of Alg. 1 is only about $28$ dB. 
In addition, for $\bar{R}$ larger than $100$ Mbps, the greedy-based algorithm cannot return a feasible solution, e.g., the return can achieve all users' demands. The percentage of SUEs/BSs which are satisfied the data demand decreases when $\bar{R}$ increases.
In these infeasible scenarios, the unsatisfied BSs/SUEs have to spend all their transmit power budgets, and cannot consume more than that amounts.
This has explained a light slow-down of the increasing trend of transmit power at $\bar{R}$ larger than $100$ Mbps for the greedy algorithm in comparison to that increasing trend due to Algorithm~\ref{alg: proposed}. 
Inversely, Algorithm~\ref{alg: proposed} has outperformed the greedy one not only in term of achieving the much lower transmission power but also satisfying users' demands in all simulated scenarios. 

\begin{figure}
    \centering
    \includegraphics[width=0.95\linewidth]{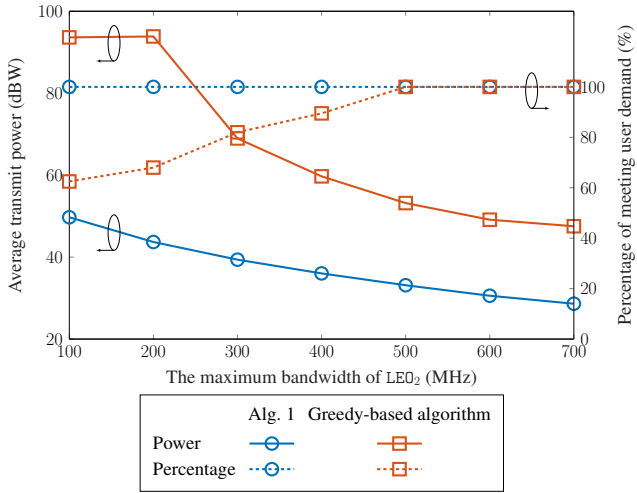}
    \caption{Average transmit power versus the maximum bandwidth of $\mathtt{LEO}_2$.}
    \label{fig:P_BW}
\end{figure}

Fig. \ref{fig:P_BW} presents the variation of transmit power versus the different bandwidth budget of $\mathtt{LEO}_2$ ($W^\LEO_2$). 
In addition, the percentages of simulated scenarios achieving feasible solution for both proposed algorithms are also provided in this figure.
As expected, SUEs and BSs consume more power if the bandwidth of $\mathtt{LEO}_2$ decreases. 
Once again, Algorithm~\ref{alg: proposed} has shown it superiority since it returns much lower transmission power than the greedy one does for all $\mathtt{LEO}_2$'s bandwidth budget.
In particular, when $W^\LEO_2$ decreases from $700$~MHz down to $100$~MHz, our first proposed algorithm helps the system consume about $28$ dBW to $49$ dBW while implementing the greedy algorithm suggests a transmit power amount of $47$ dBW to $94$ dBW.
Furthermore, Algorithm~\ref{alg: proposed} can also satisfy all users' demands in all considered scenarios while the greedy algorithm fails to do that when $W^\LEO_2$ is lower than $500$ MHz.
Hence, this figure has confirmed again the effectiveness of Algorithm~\ref{alg: proposed}.

\begin{figure}
    \centering
    \includegraphics[width=0.95\linewidth]{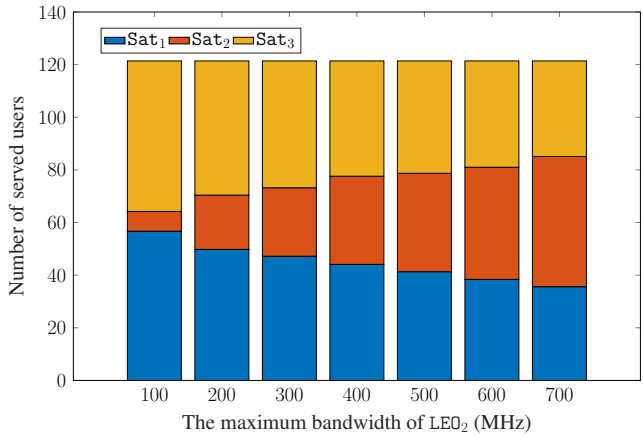}
    \caption{Number of connections of satellites versus the maximum bandwidth of $\mathtt{LEO}_2$ of a topology.}
    \label{fig:NumConnection_BW}
    \vspace{-2mm}
\end{figure}

For a more detail in the impact of the maximum bandwidth of $\mathtt{LEO}_2$, Fig. \ref{fig:NumConnection_BW} depicts the number of served users by each LEO satellite with different values of $W^\LEO_2$ according to implementing Algorithm~\ref{alg: proposed}. It can be seen that SUEs and BSs prioritize connection to $\mathtt{LEO}_2$ owing its better channel gain. When $W^\LEO_2$ decreases, the bandwidth which can be allocated for SUEs/BSs connecting to $\mathtt{LEO}_2$ has been also reduced. Hence, to optimize the transmit power and maintain the traffic demand, some of connecting SUEs/BSs have to switch their connection from $\mathtt{LEO}_2$ to the others. These results has demonstrated the flexibility of the proposed algorithm in various network circumstances.

\section{Conclusion} \label{Sec:Concl}
In this paper, we studied integrated LEO satellite and terrestrial network uplink systems, where LEO satellites serve SUEs and provide the backhaul link for cellular networks simultaneously. Subsequently, we formulated the power minimization problem including the transmit power, bandwidth allocation and LEO satellite-SUE/BS association under the user demand requirement. To solve the problem effectively, we proposed an iterative algorithm based on the alternating optimization method. Numerical results demonstrated the effectiveness of our proposed algorithm, compared with the greedy-based algorithm.

\section*{Acknowledgment}
This work has been supported by the Luxembourg National Research Fund (FNR) under the project INSTRUCT (IPBG19/14016225/INSTRUCT).

\bibliographystyle{IEEEtran}
\bibliography{Journal}
\end{document}